\documentstyle[aps,manuscript]{revtex}
\begin{document}
\draft
\title{Neutrino Flavor Oscillations Using the Dirac Equation}
\author{A. Widom and Y.N. Srivastava}
\address{Department of Physics, Northeastern University, Boston, MA 02215}
\address{and}
\address{Department of Physics \& INFN, University of Perugia, Perugia, Italy}
\date{Aug 28, 1996}
\maketitle

\begin{abstract}
The theory of neutrino flavor rotations is discussed in terms of wave
function solutions to the Dirac equation with a neutrino mass matrix. We
give a critical review of the nature of neutrino oscillations.
\end{abstract}
\pacs{PACS numbers: 1234.f}

\section{Introduction}

There is considerable interest in quantum interference patterns in space and
time which result from the notion of flavor rotations. These include Kaon
oscillations and B-meson oscillations which are thought to arise from quark
flavor rotations. Some believe that Neutrino oscillations also exist[1-4],
but presently these are merely theoretical. Such neutrino oscillations are
thought to arise from a possible flavor rotated Neutrino mass matrix. In all
such cases, the quantum interference patterns are in part due to mass
splitting, but they are also in part due to diffraction effects which would
be present even if the mass splitting were not present.

One should not forget that the invention of quantum mechanics was needed to
explain amplitude interference oscillations without any recourse to the
notion of mass splitting. For example, free electrons are described by a
Dirac wave equation%
$$
\left( -i\hbar \gamma ^\mu \partial _\mu +mc\right) \psi (x)=0,\eqno(1) 
$$
where $m$ is the electron mass. Using the usual spinors 
$$
\left( \gamma ^\mu p_\mu +mc\right) u(p)=0,\,\,\,\,u^{\dagger
}(p)u(p)=1,\,\,\,\,p^2=-(mc)^2,\eqno(1) 
$$
one may construct a solution which consists of two plane waves 
$$
\psi (x)=A_1u(p_1)e^{ip_1\cdot x/\hbar }+A_2u(p_2)e^{ip_2\cdot x/\hbar },%
\eqno(3) 
$$
and which yields an oscillating current%
$$
j^\mu (x)=ec\overline{\psi }(x)\gamma ^\mu \psi (x).\eqno(4) 
$$
It is 
$$
j^\mu (x)=ec\left( |A_1|^2\overline{u}(p_1)\gamma ^\mu u(p_1)+|A_2|^2%
\overline{u}(p_2)\gamma ^\mu u(p_2)\right) +2ec\Re e\left( A_2^{*}A_1%
\overline{u}(p_2)\gamma ^\mu u(p_1)e^{ik\cdot x}\right) ,\eqno(5) 
$$
where $\hbar k=p_1-p_2$. Note that the space-time phase interference factor
described by $\exp (ik\cdot x)$ is Lorentz invariant, and that electron
interference patterns in space and time are simply and covariantly described
by the current four vector $j^\mu (x)$. Although low energy electron
diffraction (LEED) experiments carried out every day[5,6] are most often
described in the laboratory rest frame, the description of the observed
space-time oscillation in the current $j^\mu (x)$ is not overly difficult to
describe in any Lorentz frame. Finally, the description of electron
interference by normalized wave packets (explicitly including the spin $s$), 
$$
\psi (x)=\sum_s\int A(p,s)u(p,s)e^{ip\cdot x/\hbar }d\Upsilon
,\,\,\,\,d\Upsilon =\frac{d^3{\bf p}}{\sqrt{|{\bf p}|^2+m^2c^2}}%
,\,\,\,u^{\dagger }(p,s^{\prime })u(p,s)=\delta _{s^{\prime }s},\eqno(6) 
$$
by no means eliminates the oscillations in the current 
$$
j^\mu (x)=ec\sum_{s^{\prime }s}\int \int A^{*}(p^{\prime },s^{\prime })A(p,s)%
\overline{u}(p^{\prime },s^{\prime })\gamma ^\mu u(p,s)e^{i(p-p^{\prime
})\cdot x/\hbar }d\Upsilon ^{\prime }d\Upsilon .\eqno(7) 
$$
While a sufficiently wide energy distribution in the wave packet will
produce a low oscillation visibility, LEED machines routinely produce an
electron beam which allows for the easy observation of electron diffraction
via the factor $\exp (i(p-p^{\prime })\cdot x/\hbar )$ in Eq.(7).

We have reviewed these well known features of electron diffraction
oscillations mainly because the case of massive neutrinos has for the most
part (in the literature) been treated by a completely different set of rules
than those implied by the Dirac equation. Since we feel strongly that the
Dirac equation is perfectly adequate to the task of describing freely moving
spin $1/2$ particles, including massive neutrinos, we wish to compare the
Dirac theory to those other theories which rely on a more obscure formalism.

\section{Dirac Equation}

In the Dirac theory, the massive neutrino wave functions may be denoted by $%
n_a(x)$ where $a=e,\mu ,\tau $ denotes the flavor index and $m_a$ are the
neutrino masses. The Dirac equation reads 
$$
-i\hbar \gamma ^\mu \partial _\mu \left( 
\begin{array}{c}
n_e(x) \\ 
n_\mu (x) \\ 
n_\tau (x) 
\end{array}
\right) +\left( 
\begin{array}{ccc}
m_ec & 0 & 0 \\ 
0 & m_\mu c & 0 \\ 
0 & 0 & m_\tau c 
\end{array}
\right) \left( 
\begin{array}{c}
n_e(x) \\ 
n_\mu (x) \\ 
n_\tau (x) 
\end{array}
\right) =0.\,\eqno(8) 
$$
where the wave function has twelve components, i.e. four spinor components
times three flavor components. The physical neutrino wave function $\nu (x)$
of three possible flavors is produced by the rotation%
$$
\nu (x)=\left( 
\begin{array}{c}
\nu _e(x) \\ 
\nu _\mu (x) \\ 
\nu _\tau (x) 
\end{array}
\right) =\left( 
\begin{array}{ccc}
R_{ee} & R_{e\mu } & R_{e\tau } \\ 
R_{\mu e} & R_{\mu \mu } & R_{\mu \tau } \\ 
R_{\tau e} & R_{\tau \mu } & R_{\tau \tau } 
\end{array}
\right) \left( 
\begin{array}{c}
n_e(x) \\ 
n_\mu (x) \\ 
n_\tau (x) 
\end{array}
\right) ,\eqno(9) 
$$
and obeys 
$$
\left( -i\hbar \gamma ^\mu \partial _\mu +{\cal M}c\right) \nu (x)=0,%
\eqno(10) 
$$
where the neutrino mass matrix obeys 
$$
{\cal M}_{ab}=\sum_{c=e,\mu ,\tau }R_{ac}m_cR_{cb}^{-1},\,\,\,\,\,a,b=(e,\mu
,\tau ),\eqno(11) 
$$
and the matrix $R$ is unitary 
$$
R^{\dagger }=R^{-1}.\eqno(12) 
$$
The general solution of the massive neutrino Dirac equation is best
discussed in terms of the propagator.

\section{Neutrino Propagator}

The neutrino propagator is a twelve by twelve matrix obeying 
$$
\left( -i\gamma ^\mu \partial _\mu +\frac{{\cal M}c}\hbar \right)
G(x-x^{\prime })=\delta (x-x^{\prime }),\eqno(13) 
$$
or equally well 
$$
\left( i\partial _\mu ^{\prime }G(x-x^{\prime })\gamma ^\mu +G(x-x^{\prime })%
\frac{{\cal M}c}\hbar \right) =\delta (x-x^{\prime }).\eqno(14) 
$$
Consider a space-time region $\Omega $. From the obvious identity 
$$
\nu (x)=\int_\Omega \delta (x-x^{\prime })\nu (x^{\prime })d^4x^{\prime
},\,\,\,x\in \Omega ,\eqno(15) 
$$
and Eq.(14) one finds 
$$
\nu (x)=\int_\Omega \left( i\partial _\mu ^{\prime }G(x-x^{\prime })\gamma
^\mu +G(x-x^{\prime })\frac{{\cal M}c}\hbar \right) \nu (x^{\prime
})d^4x^{\prime },\,\,\,x\in \Omega ,\eqno(16) 
$$
and 
$$
\nu (x)=\int_\Omega \left[ \left( i\partial _\mu ^{\prime }G(x-x^{\prime
})\gamma ^\mu \right) \nu (x^{\prime })+G(x-x^{\prime })\left( i\gamma ^\mu
\partial _\mu ^{\prime }\nu (x^{\prime })\right) \right] d^4x^{\prime
},\,\,\,x\in \Omega ,\eqno(17) 
$$
where Eq.(10) has been employed. Eq.(17) implies 
$$
\nu (x)=i\int_\Omega \partial _\mu ^{\prime }\left( G(x-x^{\prime })\gamma
^\mu \nu (x^{\prime })\right) d^4x^{\prime },\,\,\,x\in \Omega ,\eqno(18) 
$$
which may be converted into a ``three surface'' integral on the boundary $%
\partial \Omega $ of the region $\Omega $, i.e. 
$$
\nu (x)=i\oint_{\partial \Omega }G(x-x^{\prime })\gamma ^\mu \nu (x^{\prime
})d^3\Sigma _\mu ^{\prime },\,\,\,x\in \Omega .\eqno(19) 
$$
From Eq.(19) it is evident that the propagator allows one to compute the
full neutrino wave function for all $x\in \Omega $, if the wave function is
known on the boundary $x^{\prime }\in \partial \Omega $. Writing Eq.(19)
with flavor indices made explicit yields%
$$
\nu _a(x)=i\sum_{b=e,\mu ,\tau }\oint_{\partial \Omega }G_{ab}(x-x^{\prime
})\gamma ^\mu \nu _b(x^{\prime })d^3\Sigma _\mu ^{\prime },\,\,\,x\in \Omega
,\,\,a=(e,\mu ,\tau ),\eqno(20) 
$$
where 
$$
G_{ab}(x-x^{\prime })=\sum_{c=e,\mu ,\tau }R_{ac}R_{cb}^{-1}S(x-x^{\prime
};m_c),\eqno(21) 
$$
and $S(x-x^{\prime };m)$ is the ordinary Dirac-Feynman propagator 
$$
\left( -i\gamma ^\mu \partial _\mu +\frac{mc}\hbar \right) S(x-x^{\prime
};m)=\delta (x-x^{\prime }).\eqno(22) 
$$
The complete solution to the neutrino wave function problem is then formally 
$$
\nu _a(x)=i\sum_{b=e,\mu ,\tau }\,\sum_{c=e,\mu ,\tau }\oint_{\partial
\Omega }R_{ac}S(x-x^{\prime };m_c)\gamma ^\mu R_{cb}^{-1}\nu _b(x^{\prime
})d^3\Sigma _\mu ^{\prime },\,\,\,x\in \Omega .\eqno(23) 
$$

Eq.(13) can be solved[7] in the form 
$$
G(x-x^{\prime })=\left( i\gamma ^\mu \partial _\mu +\frac{{\cal M}c}\hbar
\right) {\cal D}(x-x^{\prime }),\eqno(24) 
$$
where%
$$
\left( -\partial _\mu \partial ^\mu +({\cal M}c/\hbar )^2\right) {\cal D}%
(x-x^{\prime })=\delta (x-x^{\prime }).\eqno(25) 
$$
From Eq.(19) and (24) it follows that 
$$
\nu (x)=\left( i\gamma ^\mu \partial _\mu +\frac{{\cal M}c}\hbar \right)
\varphi (x),\eqno(26) 
$$
where 
$$
\varphi (x)=i\oint_{\partial \Omega }{\cal D}(x-x^{\prime })\gamma ^\mu \nu
(x^{\prime })d^3\Sigma _\mu ^{\prime },\,\,\,x\in \Omega .\eqno(27) 
$$
Eq.(27) is a proper starting point for discussing possible neutrino
oscillation experiments.

\section{Outgoing Neutrino Waves}

Here we consider neutrinos (going forward in time) in a fixed Lorentz frame.
Anti-neutrinos (going backward in time) may be discussed similarly. Let $\nu
({\bf r},0)$ be the initial neutrino wave-packet function at time zero in
the Lorentz frame of interest. At later times, Eq.(27) reads (with $\beta
\equiv \gamma ^0$)%
$$
\varphi ({\bf r},t)=i\int {\cal D}({\bf r}-{\bf r}^{\prime },t)\beta \nu (%
{\bf r}^{\prime },0)d^3{\bf r}^{\prime },\,\,\,\,(t>0).\eqno(28) 
$$
From Eq.(25)%
$$
{\cal D}({\bf r}-{\bf r}^{\prime },t)=\int_{-\infty }^\infty \left( \frac{%
e^{i{\cal P}(E)|{\bf r}-{\bf r}^{\prime }|/\hbar }}{4\pi |{\bf r}-{\bf r}%
^{\prime }|}\right) e^{-iEt/\hbar }\left( \frac{dE}{2\pi \hbar c}\right) 
\eqno(29) 
$$
where the flavor matrix 
$$
c{\cal P}(E)=\sqrt{E^2-({\cal M}c^2)^2}.\eqno(30) 
$$
Putting 
$$
\varphi ({\bf r},t)=\int_{-\infty }^\infty \varphi _E({\bf r})e^{-iEt/\hbar
}\left( \frac{dE}{2\pi \hbar c}\right) ,\eqno(31) 
$$
one finds 
$$
\varphi _E({\bf r})=\int \left( \frac{ie^{i{\cal P}(E)|{\bf r}-{\bf r}%
^{\prime }|/\hbar }}{4\pi |{\bf r}-{\bf r}^{\prime }|}\right) \beta \nu (%
{\bf r}^{\prime },0)d^3{\bf r}^{\prime }.\eqno(32) 
$$
For large distances 
$$
\varphi _E({\bf r})=\left( \frac{e^{i{\cal P}(E)|{\bf r}|/\hbar }}{|{\bf r}|}%
\right) f(E,\widehat{{\bf r}}),\,\,\,\,\widehat{{\bf r}}=\frac{{\bf r}}{|%
{\bf r}|},\,\,\,\,\,|{\bf r}|>>|{\bf r}^{\prime }|,\eqno(33) 
$$
with the production amplitude (twelve components for spin and flavor) 
$$
f(E,\widehat{{\bf r}})=\left( \frac i{4\pi }\right) \int e^{-i{\cal P}(E)%
\widehat{{\bf r}}\cdot {\bf r}^{\prime }/\hbar }\beta \nu ({\bf r}^{\prime
},0)d^3{\bf r}^{\prime }.\eqno(34) 
$$
In space and time, the asymptotic outgoing wave is 
$$
\varphi ({\bf r},t)=\frac 1{|{\bf r}|}\int_{-\infty }^\infty e^{i({\cal P}%
(E)|{\bf r}|-iEt)/\hbar }f(E,\widehat{{\bf r}})\left( \frac{dE}{2\pi \hbar c}%
\right) .\eqno(35) 
$$
Employing Eq.(11), one finds with flavor indices made explicit 
$$
\varphi _a({\bf r},t)=\frac 1{|{\bf r}|}\sum_{b=e,\mu ,\tau
}\,\,\sum_{c=e,\mu ,\tau }R_{ab}R_{bc}^{-1}\int_{-\infty }^\infty
e^{i(p_b(E)|{\bf r}|-iEt)/\hbar }f_c(E,\widehat{{\bf r}})\left( \frac{dE}{%
2\pi \hbar c}\right) .\eqno(36) 
$$
where 
$$
cp_b(E)=\sqrt{E^2-(m_bc^2)^2}.\eqno(37) 
$$
Evaluating the energy integrals on the right hand side of Eq.(36) by
steepest descents, one finds the energy 
$$
\frac \partial {\partial E}\left( p_b(E)|{\bf r}|-Et\right) _{E=E_b}=0,%
\eqno(38) 
$$
yielding 
$$
\frac{|{\bf r}|}t\equiv v_b,\,\,\,E_b=\frac{m_bc^2}{\sqrt{1-(v_b/c)^2}}%
,\,\,\,p_b(E_b)=\frac{m_bv_b}{\sqrt{1-(v_b/c)^2}},\eqno(39) 
$$
so with the proper time defined by 
$$
\tau _b=t\sqrt{1-(v_b/c)^2}=-\left( \frac{p_b(E_b)|{\bf r}|-E_bt}{c^2m_b}%
\right) ,\eqno(40) 
$$
Eq.(37) reads (putting the integration variable $E=E_b+\varepsilon $), 
$$
\varphi _a({\bf r},t)\approx \frac 1{|{\bf r}|}\sum_{b=e,\mu ,\tau
}\,\,\sum_{c=e,\mu ,\tau }R_{ab}e^{-ic^2m_b\tau _b/\hbar
}R_{bc}^{-1}F_c\left( E_b,t-\frac{|{\bf r}|}{v_b},\widehat{{\bf r}}\right) .%
\eqno(41) 
$$
where 
$$
F_c(E_b,t,\widehat{{\bf r}})=\int_{-\infty }^\infty e^{-i\varepsilon t/\hbar
}f_c(E_b+\varepsilon ,\widehat{{\bf r}})\left( \frac{d\varepsilon }{2\pi
\hbar c}\right) .\eqno(42) 
$$
Eq.(41) is the central result of this work.

Eq.(41) contains the phase factors $\exp (-ic^2m_b\tau _b/\hbar )$, which
are determined by the phase velocities, and the group envelope $F_c\left(
E_b,t-(|{\bf r}|/v_b),\widehat{{\bf r}}\right) $which propagates with the
group velocity. This decomposition (phase velocity in the phase and group
velocity in the envelope) has been standard quantum mechanics for well over
half a century. See for example the standard scattering theory treatise of
Goldberger and Watson[8]. We stress this point because in the literature
some workers put the group velocity into the phase, thus forgetting why the
phrase ``phase velocity'' exists.

\section{Experimental Implications}

In experiments, there is a total outgoing amplitude (for a neutrino of
flavor $a$, when the initial neutrino had flavor $c$) given by Eq.(41); It
is 
$$
{\cal F}_{c\rightarrow a}^{total}=\left[ \sum_{b=e,\mu ,\tau
}R_{ab}e^{-ic^2m_b\tau _b/\hbar }R_{cb}^{*}F_c\left( E_b,t-\frac{|{\bf r}|}{%
v_b},\widehat{{\bf r}}\right) \right] .\eqno(43) 
$$
The absolute value squared of the amplitude is then 
$$
\left| {\cal F}_{c\rightarrow a}^{total}\right| ^2=\left[ \sum_{b=e,\mu
,\tau }\sum_{d=e,\mu ,\tau }e^{i\phi
_{bd}}R_{ab}R_{cb}^{*}R_{ad}^{*}R_{cd}F_c\left( E_b,t-\frac{|{\bf r}|}{v_b},%
\widehat{{\bf r}}\right) F_c\left( E_d,t-\frac{|{\bf r}|}{v_d},\widehat{{\bf %
r}}\right) \right] ,\eqno(44) 
$$
with the neutrino oscillation phase factors 
$$
\phi _{bd}=(m_dc^2\tau _d-m_bc^2\tau _b)/\hbar .\eqno(45) 
$$

The neutrino kinematics in the (laboratory frame) regime $m_bc^2\ll E_b$
must now be discussed. The velocities appearing in Eq.(43) are given by 
$$
v_b=c\sqrt{1-\frac{m_b^2c^4}{E_b^2}}\approx c\left( 1-\frac{m_b^2c^4}{2E_b^2}%
+\ldots \right) ,\,\,(m_bc^2\ll E_b),\eqno(46) 
$$
very slightly less than light velocity. For example, if $(m_bc^2/e)<1volt$,
and $(E_b/e)>10^6volt$, then $[(c-v_b)/c]<5\times 10^{-13}$; i.e. the
velocity (laboratory frame) is less than light velocity only by the
thirteenth significant figure. In the CHORUS and NOMAD experiments where
neutrino energies are much higher $(\sim Gev)$ than the above estimates, it
is safe to ignore the differences in the wave-packets $F_c$ in Eq.(44) and
use the phase coherent flavor conversion transition probability 
$$
{\cal P}_{coherent}(c\rightarrow a)=\sum_{b=e,\mu ,\tau }\sum_{d=e,\mu ,\tau
}R_{ab}R_{cb}^{*}R_{ad}^{*}R_{cd}\exp \left( \frac{i(m_dc^2\tau
_d-m_bc^2\tau _b)}\hbar \right) ,\eqno(47) 
$$
where Eq.(45) has been employed. Note that the use of many proper times in
Eq.(47) (i.e. the fact that each of the neutrinos with mass $m_{b,d}$ has
its own internal proper time $\tau _{b,d}$), makes the phase coherent flavor
conversion probability in Eq.(47) Lorentz invariant.

This coherent flavor conversion probability is not so obvious when neutrinos
come from distant stars. Intuitively, one would expect that the neutrino
wave packet of a heavy massive neutrino arrive on earth later than the wave
packet from a light neutrino, if the original neutrino source was (say) an
exploding star. The wave packet peaks for two different mass neutrinos would
be separated in space by $\Delta r\sim r(\Delta v/c)$ where $\Delta v$ is
the difference in the two neutrino velocities and $r$ is the distance from
the distant star to the earth. This means that 
$$
\Delta r\sim cm\left[ \frac r{light-years}\right] \left[ \frac{\Delta v}c%
\right] \times 10^{18}\,\,\,\,\,(stellar.source).\eqno(48) 
$$
Thus the phase coherence in Eq.(47) would be scrambled and one would expect
(when the source is a distant star) an incoherent flavor conversion
transition probability 
$$
{\cal P}_{incoherent}(c\rightarrow a)=\sum_{b=e,\mu ,\tau
}|R_{ab}|^2|R_{cb}|^2,\,\,\,\,\,(stellar.source).\eqno(49) 
$$
Here $\Delta r$ is so very much larger than the neutrino detecting nuclear
event length scale in the target on earth.

If the source of neutrinos is the sun, then the wave packet peak separation
is given by 
$$
\Delta r\sim cm\left[ \frac r{R_{earth-sun}}\right] \left[ \frac{\Delta v}c%
\right] \times 10^{13},\eqno(50) 
$$
where $R_{earth-sun}$ is the distance from the earth to the sun. Again the
incoherent flavor conversion transition probability applies; 
$$
{\cal P}_{incoherent}(c\rightarrow a)=\sum_{b=e,\mu ,\tau
}|R_{ab}|^2|R_{cb}|^2,\,\,\,\,\,(sun\,source).\eqno(45) 
$$
Let us now compare the central result of this work, i.e. Eq.(47), to similar
results as they appear in the literature[9-11].

\section{Previous Abuses of Dirac Notation}

The conventional textbook discussions of neutrino oscillations with flavor
mixing are simply described as follows: In Dirac notation, the three flavors
of neutrino $(\nu _e,\nu _\mu ,\nu _\tau )$ are related to those neutrinos
of fixed mass 
$$
{\cal M}|n_a>=M_a|n_a>,\,\,\,(a=e,\mu ,\tau ),\eqno(46) 
$$
by a flavor rotation matrix via 
$$
|\nu _a>=\sum_{b=e,\mu ,\tau }R_{ab}|n_b>,\,\,\,|n_a>=\sum_{b=e,\mu ,\tau
}(R^{-1})_{ab}|\nu _b>,\,\,\,(a=e,\mu ,\tau ).\eqno(47) 
$$
In the course of time, the state of a given flavor neutrino changes via 
$$
\exp \left( -iHt/\hbar \right) |\nu _c>=\sum_{b=e,\mu ,\tau }R_{cb}\exp
(-iE_bt/\hbar )|n_b>,\eqno(48) 
$$
where 
$$
E_b=\sqrt{c^4m_b^2+c^2|{\bf p}|^2},\,\,\,\,(b=e,\mu ,\tau ).\eqno(49) 
$$
and where ${\bf p}$ is the neutrino three-momentum. Eqs.(47) and (48) imply
that 
$$
<\nu _a|\exp \left( -iHt/\hbar \right) |\nu _c>=\sum_{b=e,\mu ,\tau
}R_{cb}(R^{-1})_{ba}\exp (-iE_b({\bf p})t/\hbar ).\eqno(50) 
$$
One then concludes that the transition probabilities between neutrinos of
different flavors, 
$$
P_{WRONG}(c\rightarrow a,t)=|<\nu _a|\exp \left( -iHt/\hbar \right) |\nu
_c>|^2\,????,\eqno(51) 
$$
oscillate in time. The above described conventional arguments are clear,
elegant, appealing, and wrong.

One may suspect that the conventional discussion above has some problems
with Lorentz invariance by applying it (as very many workers have done) to a
neutrino produced in the sun and detected on earth. Suppose that a nucleus
``at rest'' in the sun fires off a beta-decay neutrino. There will be
according to the above (incorrect but conventional) discussion a neutrino
superposition of three {\em different energies} $E_b$, ($b=e,\mu ,\tau $)
but {\em all with the same three-momenta} ${\bf p}$. Not all of the nuclei
on the sun are ``at rest''. Now suppose that a moving nucleus fires off a
beta-decay neutrino. Can one believe that the three energies are yet again
all different but the three-momenta are all yet again the same? The answer
is obviously no! If Lorentz symmetry is invoked, then momentum and energy
together form a four vector $p=({\bf p},E/c)$. One should put $%
p_b^2=-(m_bc)^2$ for an on mass shell neutrino, and which of the four
momentum {\em components} may be the same or may be different will then
depend on the reference frame. At least the {\em possibility} that the
neutrino is in a superposition of different three-momenta should be {\em %
considered}. This requires that Eq.(49) for an on mass shell neutrino should
be replaced by {\em \ } 
$$
E_b=\sqrt{c^4m_b^2+c^2|{\bf p}_b|^2},\,\,\,\,(b=e,\mu ,\tau ).\eqno(52) 
$$
With this {\em replacement} the conventional discussion is {\em now} clear,
elegant, appealing, and {\em still }wrong.

Exactly what is the amplitude $<\nu _{final}|\exp (-iHt/\hbar )|\nu
_{initial}>$ supposed to mean? If we assume that the neutrino is a simple
spin $1/2$ particle without further internal structure (beyond the flavor
label), then perhaps it is supposed to mean something like 
$$
<\nu _{final}|\exp (-iHt/\hbar )|\nu _{initial}>=\int \nu _{final}^{\dagger
}({\bf r})\exp (-iHt/\hbar )\nu _{initial}({\bf r})d^3r\,\,\,\,????\eqno(53) 
$$
We have chosen a time $t$ (which means we have chosen a Lorentz reference
frame), and a standard Dirac Hamitonian (with flavor indices implicit for
example in the mass matrix ${\cal M}$) 
$$
H=c{\bf \alpha \cdot p}+{\cal M}c^2\beta ,\,\,\,\,\,\,\,\,\,({\bf p}=-i\hbar 
{\bf \nabla }),\eqno(54) 
$$
If the above is indeed the case, then there is a remaining problem. In
Eq.(53), the inner product contains an {\em integral over space}.

Space is not the same as time. When there are energy superpositions there
will be time oscillations. When there are momentum superpositions there will
be space oscillations. Furthermore, when the neutrino comes from the sun to
the earth, we know mainly {\em where} the neutrino was detected. The
predicted oscillations should be (at least in part) in space. An {\em %
integration over space} seems a peculiar way to get quantum interference
(oscillation) diffraction patterns {\em that exist in space}. Consider
electron diffraction (in real experiments scattering electrons off a crystal
face). If ``thought experiments'' are more appealing, then consider two-slit
electron diffraction. To get the interference (oscillation) pattern in
space, one absolute squares a spatial amplitude. If you integrate over
space, then you get a total probability of unity (which is true but not
really of much use in discussing diffraction oscillations).

Our suggestion has been in the previous sections the following: (i) If one
wants to study neutrino oscillations in space-time, then it is best to study
a wave function which actually depends on space and time; e.g. try the Dirac
equation Eq.(10). (ii) If one wants to study how the neutrinos are
distributed in space and time, then it is best to study the current $j^\mu
(x)=c\overline{\nu }(x)\gamma ^\mu \nu (x)$\thinspace which actually
describes the distribution in space and time. (iii) Finally, if one wants to
study how a neutrino can propagate from the sun to the earth (or across a
large laboratory), then it is best to study the propagator Eqs.(13) and (19)
which actually describes how the neutrino propagates.

\section{Conclusion}

The coherent neutrino oscillation flavor conversion probability

$$
{\cal P}_{coherent}(c\rightarrow a)=\sum_{b=e,\mu ,\tau }\sum_{d=e,\mu ,\tau
}R_{ab}R_{cb}^{*}R_{ad}^{*}R_{cd}\exp \left( \frac{i(m_dc^2\tau
_d-m_bc^2\tau _b)}\hbar \right) \eqno(55) 
$$
depends on the proper times of the massive neutrinos; i.e. 
$$
\hbar \phi _b=-m_bc^2\tau _b=(p_b|{\bf r}|-E_bt).\eqno(56) 
$$
Some previous workers employ only the energy-time factor $E_bt$ in the phase 
$\phi _b$, while other previous workers employ only the momentum-space
factor $p_b|{\bf r}|$ in the phase $\phi _b$. To consider only energy-time
or momentum-space and (not both) tells only half the story. The true Lorentz
invariant phase $\phi _b$\thinspace requires both.

Finally, neutrino flavor conversions may be described sometimes by ${\cal P}%
_{coherent}(c\rightarrow a)$ and sometimes by ${\cal P}_{incoherent}(c%
\rightarrow a)$, depending upon what is assumed about neutrino masses,
rotation matrix elements, energies of neutrino sources and length scales of
actual experiments. There is no universal simple formula which holds true in
all regimes. \vfill \eject

\centerline {\bf REFERENCES}

\noindent
[1] S. M. Bilenky and S. T. Petcov, {\it Rev. Mod. Phys. }{\bf 59} 671
(1987);

\noindent
[2] F. Boehm and P. Vogel, {\it Physics of Massive Neutrinos}, Cambridge
University Press, Cambridge, (1987).

\noindent
[3] R. E. Shrock, {\it Phys. Rev. D }{\bf 24, }1232 (1981); {\bf 24}, 1275
(1981).

\noindent
[4] J. N. Bahcall, {\it Neutrino Physics and Astrophysics}, Cambridge
University Press, Cambridge, (1989).

\noindent
[5] P. J. Estrup, {\it LEED Studies of Surface Layers}, in {\it %
Characterization of Metal and Polymer Surfaces}, Academic Press, New York
(1977).

\noindent
[6] D. B. Williams, {\it Practical Electron Microscopy in Materials Science}%
, Verlag Chemie, Weinheim (1984).

\noindent
[7] J. Schwinger, {\it Particles, Sources and Fields } Vol.I, Addison-Wesley
Publishing Company Inc., Vol. I, Reading (1989).

\noindent
[8] M. L. Goldberger and K. M. Watson, {\it Collision Theory } p.61, John
Wiley\& Sons, New York, (1965).

\noindent
[9] R. G. Winter, {\it Lett. Nuovo Cimento }{\bf 30 }(1981) 101.

\noindent
[10] B. Kayser, {\it Phys. Rev. D }{\bf 24}, 110 (1981).

\noindent
[11] C. Giunti, C. W. Kim, J. A. Lee, and U. W. Lee, {\it Phys. Rev. D }{\bf %
48}, 4310 (1993).

\end{document}